\documentclass[12pt]{article}
\usepackage{epsfig}
\begin{document}
\title{ Pad\'e approximation of the $S$-matrix as a way of locating
quantum resonances and bound states}
\author{
{\bf S. A. Rakityansky$^{1,2}$, S. A. Sofianos$^{3}$
and N. Elander$^{2}$}\\
\begin{tabular}{ll}
1)& {\small\it Dept. of Physics, University of Pretoria, Lynnwood
  Road, Pretoria 0002, South Africa}\\
2)& {\small\it Dept. of Physics,
Stockholm University, Alba Nova University Center,}\\
&{\small\it Stockholm, SE-106 91, Sweden}\\
3)& {\small\it Dept. of Physics, UNISA, 
Pretoria 0003, South Africa}
\end{tabular}
}
\maketitle
\begin{abstract}
\noindent
It is shown that the spectral points (bound states and resonances)
generated by a central potential of a single-channel problem, can be
found using rational parametrization of the $S$-matrix. To achieve
this, one only needs values of the $S$-matrix along the real positive
energy axis. No calculations of the $S$-matrix at complex energies or
a complex rotation are necessary. The proposed method is therefore
universal in that it is applicable to any potential (local, non-local,
discontinuous, etc.) provided that there is a way of obtaining the
$S$-matrix (or scattering phase-shifts) at real collision energies.
Besides this, combined with any method that extracts the phase-shifts
from the scattering data, the proposed rational parametrization
technique would be able to do the spectral analysis using the
experimental data.
\end{abstract}
\vspace{.5cm}
\noindent
PACS number(s): {03.65.Nk, 03.65.Ge, 24.30.Gd}\\[5mm]
Published in:\quad
{\large\bf J. Phys. A: Math. Theor. 40 (2007) 14857-14869}\\[5mm]

\section{Introduction}

A full understanding of the properties of a quantum system and prediction
of its behaviour cannot be achieved without knowing its spectrum,
i.e. the energies of its bound states and resonances. The problems of
this kind emerge not only in fundamental research concerning
particles, nuclei, and atoms, but also in engineering. For example, modern
semiconductor devices based on nano-structures, cannot be properly
designed without accurate treatment of various transitions among
stationary and quasi-stationary states of the charge carriers (see,
for example, Ref.\cite{nano2} and references therein).

The problem of locating bound states is as old as quantum mechanics
itself. Over a century, plenty of exact and approximate methods were
developed for solving it. The notion of quasi-stationary states (or
quantum resonances) emerged at later stages of the development of
quantum mechanics.  The attention to such states was drawn by Gamow in
his pioneering works on the $\alpha$-decay \cite{gamow}.

At the beginning, the development of the methods for locating
resonances was hindered by computational difficulties, which are more
challenging than that of the bound state problem. The progress in this
field was therefore delayed till the advent of modern computers.
Another consequence of these difficulties was that the methods for
solving the bound and quasi-bound state problems were developed
separately despite the fact that the bound and resonant states have
essentially the same mathematical nature. Indeed, they correspond to
the $S$-matrix poles on the complex energy surface at all of which the
asymptotics of the solutions of the Schr\"odinger equation has the
same functional form, namely, the outgoing wave. The only difference
between bound and quasi-bound states is that the corresponding poles
are situated in different domains of the energy surface. As a result,
the wave function asymptotics look deceptively different. Being
found at real negative energies, the bound state solutions at large
distances exponentially decay. However, the decaying exponential
function is the same outgoing wave, but taken with pure imaginary
momentum that corresponds to a negative energy.

Another fact that makes the kinship between the bound and
quasi-bound states apparent, is their mutual transformations. This
happens when one increases or decreases depth of the potential. A
gradually deepening potential ``sucks'' the resonances in. The
corresponding poles of the $S$-matrix move towards the threshold
point ($E=0$) and eventually cross it over to the bound state
domain.

Common mathematical ground of the bound and resonant states implies
that it should be a unified way of locating them. In quest for such a
way most of the modern methods were developed. A review of the existing
methods for solving the quasi-bound state problem can be found, for
example, in Ref.\cite{kukulin}. One of the new unified approaches not
mentioned in that book, is based on a direct calculation of the
Jost-function at complex energies \cite{exactmethod} (for references
concerning recent development of this method, see also
Ref.\cite{nano1}).

All the methods for locating resonances can be divided in two
groups. One of them comprises the approaches, in which all the
calculations are done only at real energies. Such methods are rather
simple but have limited abilities. They usually fail for wide as well
as extremely narrow resonances and are not unified in the
abovementioned sense. The other group combines the methods based on
locating the $S$-matrix poles (or Jost function zeros) on the complex
energy surface. These approaches are very powerful, accurate, and
unified. The price one has to pay for that is that they require rather
sophisticated calculations.

Even the unified complex-energy methods are not fully universal. They
are not applicable, for example, to potentials that cannot be
analytically continued to complex values of the distance $r$. They
have difficulties in dealing with energy-dependent and non-local
potentials. The methods based on the complex rotation, cannot reach
the so-called virtual states that correspond to the $S$-matrix poles
at negative imaginary momenta, i.e. at negative energies on the
unphysical sheet of complex energy surface.

Thus, one can choose either a simple but inaccurate real-energy method
or a powerful but complicated complex-energy one. None of them,
however, is universal. It would be desirable to have a method that
could combine in itself both the simplicity of the real-energy
approaches and the power and accuracy of the complex-energy ones.  In
this paper, we suggest such a method. Besides this, the proposed
method is insensitive to the nature of the interaction potential and
therefore can be considered as universal.

The idea is based on the fact that two analytic functions
coinciding on a curve segment, are identical everywhere in the
complex plane (the so-called coincidence
principle\cite{complex_analysis}).  Therefore if we calculate
(using any appropriate or available method) the $S$-matrix along
the real positive axis of the energy plane and accurately fit the
obtained values with a meromorphic function, for example, with a
Pad\'e function, then we can expect that this approximate function
will have practically the same singularities at complex energies
as the exact $S$-matrix. In this way, we can locate the spectral
points (bound, virtual, and resonance states) by locating these
singularities. In the suggested method, all the poles of the
Pad\'e function are located at once, because their coordinates are
the fitting parameters and thus they are determined within the
fitting procedure.

By constructing the Pad\'e approximant, we actually do an analytic
continuation of the $S$-matrix from a segment of the real axis,
where it is given at a number of discrete points, to complex
energies. Similar (but different) procedures in which function
values on the complex plane are obtained using the knowledge of
its values at a discrete set of real points, were more than once
used before and proved to be successful. As an example, we can
mention Ref. \cite{massen} where a potential $V(r)$ numerically
calculated at a set of real points $r_1, r_2, \dots, r_N$ was used
within a complex rotation method for locating resonances. Another
example is the Pad\'e approximation of the Titchmarsh-Weyl
$m$-function that was used in Ref. \cite{weyl} also for locating
resonances.

\section{Rational approximation of the $S$-matrix}

Let us assume that we know the complex-valued function $S_\ell(k)$ for
all real values of the collision momentum
$k\in[0,\infty)$. Alternatively, we can have the real phase-shift
function $\delta_\ell(k)$ related to the $S$-matrix as
\begin{equation}
\label{Sdelta}
               S_\ell(k)=\exp\left[i2\delta_\ell(k)\right]\ .
\end{equation}
In practical calculations the interval $[0,\infty)$ is reduced, of
course, to a finite segment $[k_{\rm min},k_{\rm max}]$ with
sufficiently small $k_{\rm min}$ and large $k_{\rm max}$, that covers
all significant oscillations of the function $\delta_\ell(k)$.

We are going to find an approximate function $\tilde S_\ell(k)$ such
that the difference $S_\ell(k)-\tilde S_\ell(k)$ is minimal on the
segment $[k_{\rm min},k_{\rm max}]$. Keeping in mind the well-known
fact (see, for example, Ref.\cite{complex_analysis}) that two analytic
functions are identical everywhere if they coincide on a continuous
segment, we then expect that the approximate $S$-matrix $\tilde
S_\ell(k)$ will have almost the same singularities (resonance and
bound state poles) as the exact one.  The more accurately we
approximate the $S$-matrix on the real axis, the less different will
be the poles of $\tilde S_\ell(k)$ from the corresponding poles of
$S_\ell(k)$.

There are many ways of approximating $S_\ell(k)$.  A choice for the
functional form of $\tilde S_\ell(k)$ is determined by the fact that
it should have simple poles, i.e. at certain points $k_i$ be
proportional to $\sim (k-k_i)^{-1}$. Such behaviour is provided by a
ratio of two polynomials
\begin{equation}
\label{ratioPP}
               \tilde S_\ell(k)=
               \frac{a_0+a_1k+a_2k^2+\dots+a_Mk^M}
                    {b_0+b_1k+b_2k^2+\dots+b_Nk^N}\ ,
\end{equation}
which in numerical analysis is known as the Pad\'e approximation
of the order $[N,M]$ when the parameters $a_m$ and $b_n$ are
chosen to reproduce the exact function and all its derivatives up
to the order $M+N$ at $k=0$ (see, for example, Refs.
\cite{num_recipes, baker_gammel}).

Here, we will use a special form of the approximation (\ref{ratioPP}).
Firstly, we observe that at high energies the $S$-matrix tends to
unity \cite{Newtonbook}, which can only be achieved if both the
numerator and denominator polynomials have the same degree,
i.e. $M=N$, and $a_M=b_N$. Secondly, at zero energy the $S$-matrix is
unity as well \cite{Newtonbook}, which implies that $a_0=b_0$ (without
loosing the generality, we can assume that $a_0=b_0=1$).Thirdly, the
fitting parameters cannot in our case be found using the $S$-matrix
derivatives, which are not available. Instead, we use the algorithm
described next.

\section{Fitting parameters}
To begin with, we re-arrange the polynomials in the form that includes
their zeros explicitly, namely,
\begin{equation}
\label{productPP}
               \tilde S_\ell(E)=\prod_{n=1}^{N}
               \frac{k-\alpha_n}{k-\beta_n}\ .
\end{equation}
Knowing some general properties of the $S$-matrix, we can simplify the
expression (\ref{productPP}). This can be done as follows.  First of
all, we notice that the exact $S$-matrix is the following ratio
\begin{equation}
\label{ratioFF}
               S_\ell(k)=
               \frac{f_\ell^{(\rm out)}(k)}{f_\ell^{(\rm in)}(k)}\ ,
\end{equation}
where $f_\ell^{(\rm in/out)}(k)$ are the Jost
functions\cite{Taylorbook} that are the amplitudes of the incoming and
outgoing waves in the asymptotics or the radial wave function,
\begin{equation}
\label{uass}
               u_\ell(k,r)\ \mathop{\longrightarrow}_{r\to\infty}
               f_\ell^{(\rm in)}(k)h_\ell^{(-)}(kr)+
               f_\ell^{(\rm out)}(k)h_\ell^{(+)}(kr)\ ,
\end{equation}
with the incoming and outgoing spherical waves being represented by
the corresponding Riccati-Hankel functions $h_\ell^{(-)}(kr)$ and
$h_\ell^{(+)}(kr)$.

\begin{figure}[ht!]
\centerline{\epsfig{file=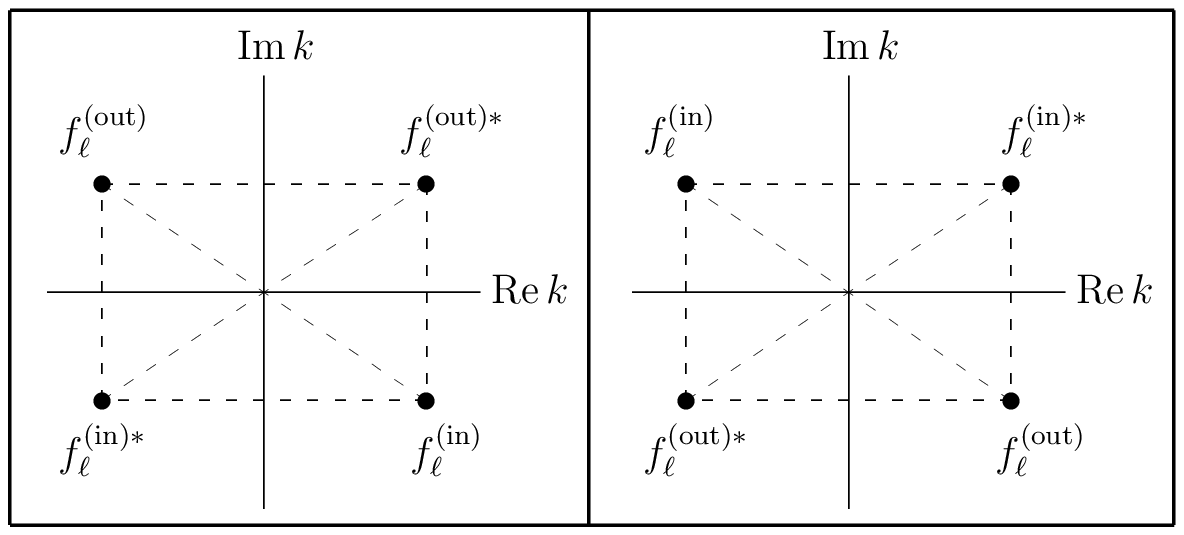}}
\caption{\sf
Symmetry properties of the Jost functions on the $k$-plane.
The dashed lines connect the points at which the values indicated next
to them are identical.
}
\label{fig.ksymmetry}
\end{figure}

Now, we use the fact that $f_\ell^{(\rm in)}(k)$ and
$f_\ell^{(\rm out)}(k)$ are not independent. Indeed, the
incoming and outgoing waves swap their roles when the momentum
$k$ changes its sign and also under the operation of complex
conjugation. It is not difficult to show (see, for example,
Ref. \cite{Taylorbook}) that this implies the symmetry properties
summarized in Fig. \ref{fig.ksymmetry}.
In particular, we have
\begin{equation}
\label{finfoutm}
          f_\ell^{(\rm out)}(k)=f_\ell^{(\rm in)}(-k)\ ,
\end{equation}
and thus
\begin{equation}
\label{ratioFFin}
               S_\ell(k)=
               \frac{f_\ell^{(\rm in)}(-k)}{f_\ell^{(\rm in)}(k)}\ .
\end{equation}
This means that not all parameters in Eq. (\ref{productPP}) are
independent. The numerator polynomial should be the same as the
denominator one but taken with negative $k$. Therefore
\begin{equation}
\label{productB}
               \tilde S_\ell(k)=(-1)^N\prod_{n=1}^{N}
               \frac{k+\beta_n}{k-\beta_n}\ .
\end{equation}
As compared to Eq. (\ref{productPP}), this not only reduces the
number of fitting parameters in half, but also improves the quality of
the approximation since the correct structure of the $S$-matrix, given
by Eq. (\ref{ratioFFin}), is taken into account.

The procedure of finding the parameters $\beta_n$ consists of two
stages. At the first stage, we once again re-arrange the polynomials
in Eq. (\ref{productB}) to the form (\ref{ratioPP}),
\begin{equation}
\label{ratioKK}
               \tilde S_\ell(k)=
               \frac{\displaystyle 1+\sum_{n=1}^Na_nk^n}
                    {\displaystyle 1+\sum_{n=1}^N(-1)^na_nk^n}\ ,
\end{equation}
where the numerator and denominator are divided by the product
$\beta_1\beta_2\cdots\beta_N$, which gives $a_0=b_0=1$. In order to
find the parameters $a_n$, we multiply Eq. (\ref{ratioKK}) by the
denominator of its right hand side and re-write it as
\begin{equation}
\label{system_0}
     \sum_{n=1}^N\left[1+(-1)^{n+1}\tilde S_\ell(k)\right]k^na_n
     = \tilde S_\ell(k)-1\ .
\end{equation}
The last equation, taken at $N$ different points $k_1, k_2,\dots,
k_N$, constitutes a system of linear equations
\begin{equation}
\label{system_1}
     \sum_{n=1}^NA_{mn}a_n = B_m\ ,\qquad m=1,2,\dots,N
\end{equation}
determining the unknown coefficients $a_n$. The matrices of
this system are defined as
\begin{equation}
\label{system_A}
     A_{mn}=\left[1+(-1)^{n+1}S_\ell(k_m)\right]k_m^n
\end{equation}
and
\begin{equation}
\label{system_B}
     B_{m}=S_\ell(k_m)-1
\end{equation}
with $S_\ell(k_m)$ being known values of the $S$-matrix on the
interval $[k_{\rm min},k_{\rm max}]$. Therefore, by solving matrix
equation (\ref{system_1}), we can determine the parameters $a_n$.

What we actually need are the $S$-matrix poles, i.e. the parameters
$\beta_n$. So, at the second stage, the parameters $\beta_n$ are
determined as the complex roots of the polynomial
\begin{equation}
\label{polynomial}
      P_N(k)=1+\sum_{n=1}^N(-1)^na_nk^n\ .
\end{equation}
There are many robust and fast algorithms for finding such roots (see,
for example, Ref. \cite{num_recipes}).

\section{Meaningful and spurious poles}
What are the mathematical and physical meaning of the poles of the
approximate $S$-matrix $\tilde{S}_\ell(k)$? As we know, all the poles
of the exact $S$-matrix are the spectral points, i.e. the bound and
virtual staes (if any) as well as the resonances. Is this also valid
for the poles of $\tilde{S}_\ell(k)$?

At a first sight, it seems that we should give a positive answer to
this question. Indeed, the functions ${S}_\ell(k)$ and
$\tilde{S}_\ell(k)$ coincide on a segment of the real axis and thus,
according to the coincidence principle of the complex analysis, they
should be identical everywhere on the complex $k$-plane.
There are, however, two facts that cast some doubts on this reasoning.

Firstly, the function $\tilde{S}_\ell(k)$ coincides with ${S}_\ell(k)$
not on a continuous segment, but only at a finite number of discrete
points $k_1,k_2,\dots,k_N$. And secondly, the number $N$ of these
points and thus the number of poles that the function (\ref{productB})
may have, is chosen arbitrarily.

It is natural to expect that the smaller is the difference between
$\tilde{S}_\ell(k)$ and ${S}_\ell(k)$ on the chosen segment, the
better is the approximation of the $S$-matrix at complex momenta.
This difference can be made smaller by increasing the number $N$ of
the points, in other words, the number of poles of
$\tilde{S}_\ell(k)$. Surely, not all of these poles have physical
meaning. Some of them may appear at ``wrong'' places, because of the
non-zero difference $\tilde{S}_\ell(k)-{S}_\ell(k)$ in between the
points $k_1,k_2,\dots,k_N$. On the other hand, some of them must be
close to the ``true'' poles, otherwise the approximate $S$-matrix
would be too different from the exact one.

Those poles of $\tilde{S}_\ell(k)$ that are close to the ``true''
$S$-matrix poles are meaningful, while all the other poles are
spurious. It is obvious that the positions of the meaningful poles are
``tied'' to the the ``true'' poles, while the spurious poles are
``free to move'' and appear randomly depending on the choice of the
fitting points $k_1,k_2,\dots,k_N$.

Therefore, there is a simple way to distinguish meaningful poles from
the spurious ones. Indeed, repeating the calculations with different
number of fitting points, we can easily find those poles that appear
more or less at the same positions. All the other poles should be
regarded as spurious. In this way, we can also find an appropriate
number $N$, with which the meaningful poles converge to the ``true''
poles within a required accuracy.

When doing the calculations, we found (see the next section) that the
spurious poles in many cases appear in symmetric pairs that almost
exactly cancel each other in the product (\ref{productB}). For
example, if a spurious pole appears at $k=\beta_n$ then it is
accompanied by another pole at $k=\beta_{m}$ such that $\beta_m\approx
-\beta_n$. As a result, in the product (\ref{productB}), we have the
factor
\begin{equation}
\label{spurious}
                  \frac{(k+\beta_n)(k+\beta_m)}
                  {(k-\beta_n)(k-\beta_m)}\approx 1\ ,
\end{equation}
which is practically unity everywhere except the immediate vicinity of
the spurious poles.

\section{Numerical examples}
The basic idea of the proposed method rests on a rigorous mathematical
fact of the identity of two functions that coincide on a continuous
curve segment. In numerical calculations, however, we can only
guarantee that the exact and approximate functions $S_\ell(k)$ and
$\tilde{S}_\ell(k)$ coincide at a number of discrete points along the
real $k$-axis. Of course, after the parameters of $\tilde{S}_\ell(k)$
are fixed, we can always check how this approximate function
reproduces the exact one at the intermediate points. Our hope is based
on the well-known fact that rational interpolation of the type
(\ref{ratioPP}) works very well even with just few matching
points\cite{num_recipes}. Certainly, the accuracy of the approximation
is best near the matching points and should deteriorate when we move far
away into the complex plane.

In order to test how this rational interpolation works in our method,
we performed calculations for several well-studied potentials, whose
spectral points are known.  Since in the proposed method nothing
special is associated with the angular momentum, we tested here only
the case $\ell=0$. The exact values of the $S$-matrix at the fitting
points were calculated using the Jost function method described in
Ref. \cite{exactmethod}. The same method was used to locate the
the exact spectral points, with which the approximate $S$-matrix poles
were compared.

The first of the testing potentials is an exponential well,
\begin{equation}
\label{pot.NN1}
       V_{NN}^{(1)}(r)=-W^{(1)}\exp\left(-r/R_1\right)
\end{equation}
with $W^{(1)}=154.06$\,MeV and $R_1=0.76$\,fm, which roughly
describes the $S$-wave proton-neutron interaction in the triplet state
(total spin = 1). With $\hbar^2/(2m)=41.47\,{\rm MeV\cdot fm^2}$ this
potential generates a bound state (the deuteron) at
$E=-2.2244674752$\,MeV. With $\ell=0$, it does not generate any other
poles of the $S$-matrix (at least within any physically reasonable
domain of the complex plane).

Since there is only one ``true'' pole, we expect that it is not
necessary to have many fitting points in order to reproduce this pole
using the Pad\'e approximation. And indeed, starting with just one
point, we see a very rapid convergence, so that at $N=10$ the exact
value of the binding energy is reproduced to eight decimal places,
namely, we obtain $-2.2244674718$\,MeV. As the fitting segment for
these calculations, we (arbitrarily) chose the interval from
$E_1=1\,{\rm MeV}$ to $E_N=10\,{\rm MeV}$ on the real axis. The
fitting points,
\begin{equation}
\label{e1e2dotsen}
    E_n=E_1+\frac{E_N-E_1}{N-1}(n-1)\ ,\qquad n=1,2,\dots,N\ ,
\end{equation}
were evenly distributed over this interval and the corresponding
momenta were taken as $k_n=\sqrt{2mE_n/\hbar^2}$.  Table
\ref{table.nn1} shows the meaningful as well as all spurious poles up
to $N=5$. It is amazing, but even with only one fitting point we
already get a reasonable value for the binding energy.

The second testing potential also describes the $S$-wave
proton-neutron interaction, but in the singlet state (total spin = 0),
\begin{equation}
\label{pot.NN0}
       V_{NN}^{(0)}(r)=-W^{(0)}\exp\left(-r/R_0\right)
\end{equation}
with $W^{(0)}=104.20$\,MeV and $R_0=0.73$\,fm. It has a weaker
attraction and thus, instead of a bound state, it generates a virtual
state at $E=-0.0660644719$\,MeV. Similarly to the triplet case, this
is the only spectral point in the physically reasonable domain. The
convergence here is even more faster (see Table \ref{table.nn0}, where
we show the results up to $N=5$).  With the same choice of the fitting
interval and fitting points as we used for the first potential, ten
decimal places of the virtual state energy are reproduced already at
$N=7$.

\begin{figure}[ht!]
\centerline{\epsfig{file=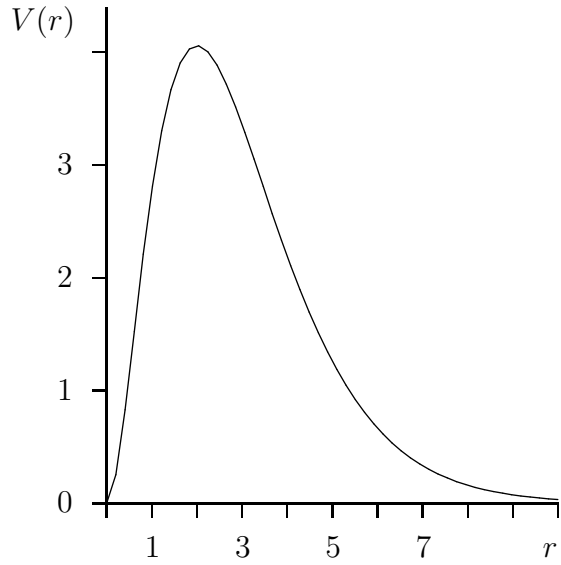}}
\caption{\sf
Testing potential (\protect\ref{pot1.formula}) in the arbitrary units such
that $\hbar=m=1$.
}
\label{fig.pot1}
\end{figure}

The next example is an exponential hump, shown in
Fig. \ref{fig.pot1}. It is positive everywhere and therefore can only
support resonant states. In the units such that $\hbar=m=1$, the
functional form of this potential is
\begin{equation}
\label{pot1.formula}
             V(r)=7.5r^2\exp(-r)\ .
\end{equation}
The exact $S$-matrix corresponding to this potential, has an infinite
number of poles forming a string that goes down the $k$-plane to
infinity. The exact locations of the first nine of these poles are
given in Table \ref{table.V1exact}. These values of the resonance
energies were obtained using a very accurate method, which is based on
a combination of the complex rotation and a direct calculation of the
Jost function (see Ref. \cite{exactmethod}).

\begin{figure}[ht!]
\centerline{\epsfig{file=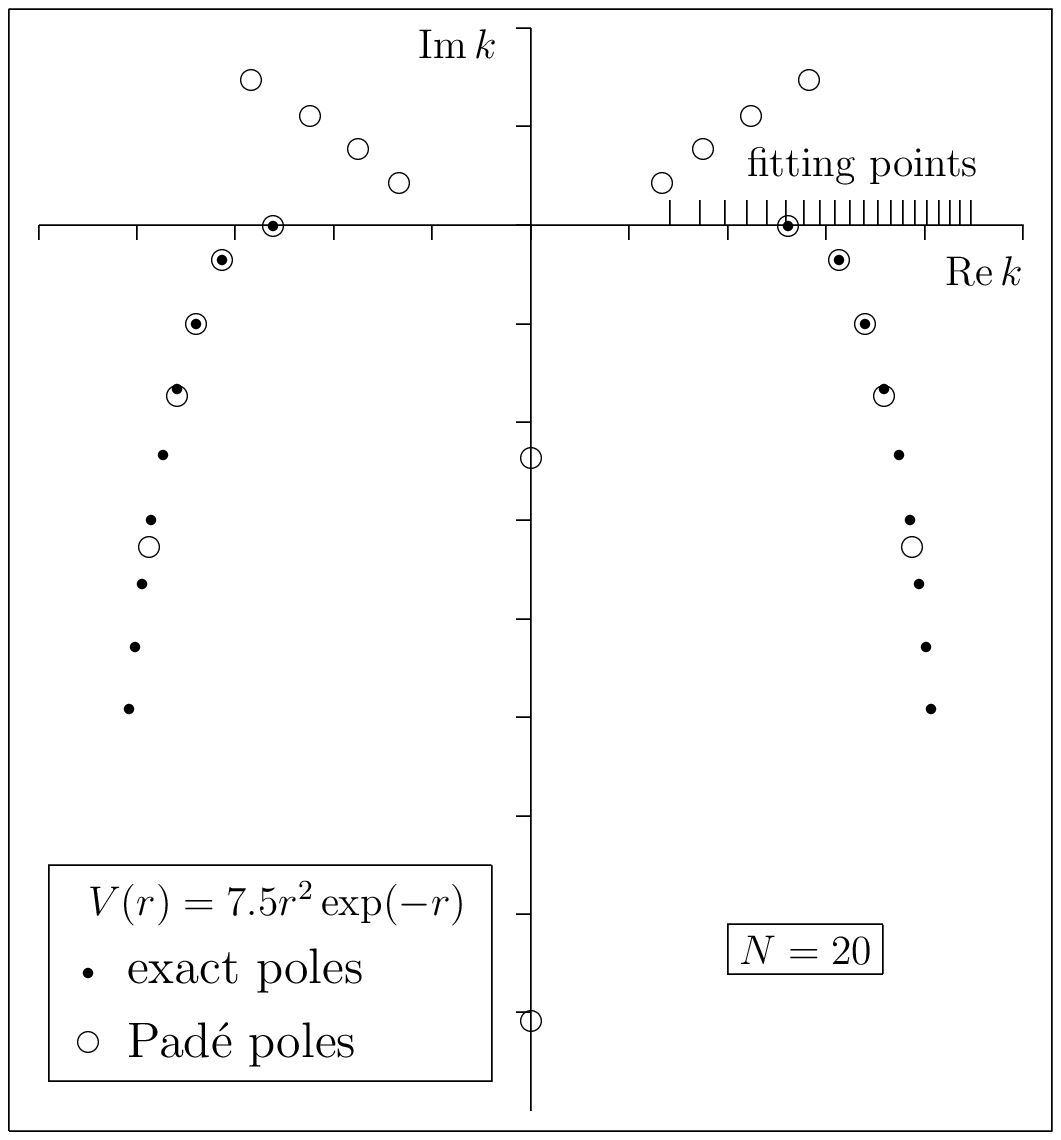}}
\caption{\sf
The exact positions of the $S$-wave resonance poles (dots) on the momentum
plane for the potential (\protect\ref{pot1.formula}), and the
corresponding poles of the Pad\'e approximation (open circles)
obtained with 20 fitting points evenly distributed over the energy
interval $1\le E\le 10$ on the real $E$-axis.
The corresponding fitting points on the $k$-axis are indicated by
vertical bars.
}
\label{fig.v1n20}
\end{figure}

\begin{figure}[ht!]
\centerline{\epsfig{file=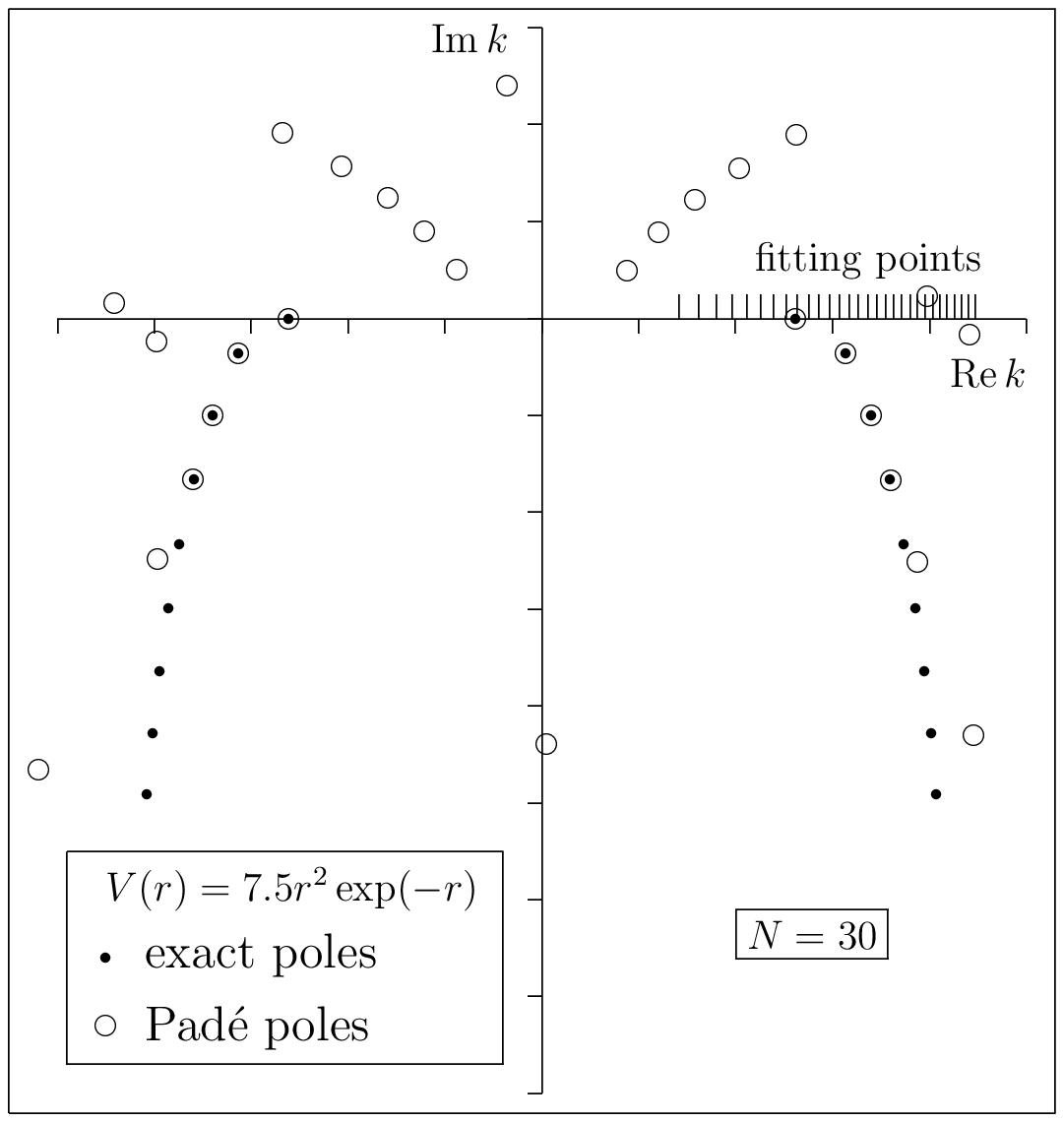}}
\caption{\sf
The exact positions of the $S$-wave resonance poles (dots) on the momentum
plane for the potential (\protect\ref{pot1.formula}), and the
corresponding poles of the Pad\'e approximation (open circles)
obtained with 30 fitting points evenly distributed over the energy
interval $1\le E\le 10$ on the real $E$-axis.
The corresponding fitting points on the $k$-axis are indicated by
vertical bars.
}
\label{fig.v1n30}
\end{figure}

It is naturally to expect that the approximate $S$-matrix will
reproduce the beginning of the string of resonance poles, i.e. the
most significant poles that are close to the real axis and thus to the
fitting segment. This is indeed the case as is seen in
Figs. \ref{fig.v1n20} and \ref{fig.v1n30}.  The first of these two
figures shows the distribution of the exact and approximate $S$-matrix
poles over the $k$-plane, with the number of fitting points
$N=20$. The second figure shows the same, but for $N=30$ (few spurious
poles that are too far away from the origin are not shown). In both
cases the fitting points were uniformly placed on the interval $1\le
E\le 10$, as is given by Eq. (\ref{e1e2dotsen}).  In
Figs. \ref{fig.v1n20} and \ref{fig.v1n30}, the corresponding points on
the real $k$-axis are indicated by vertical bars. As is seen, the
distances between neighboring bars are decreasing towards the right
end of the interval. This is because the relation between $k$ and $E$
is not linear.

The poles of the exact $S$-matrix are symmetric on the $k$-plane
relative to the imaginary axis (see Fig. \ref{fig.ksymmetry}). This is
a consequence of the Schwartz reflection principle and of the fact
that the incoming and outgoing spherical waves swap their roles under
the operation of complex conjugation (for proof see, for example,
Ref. \cite{Taylorbook}).  The distribution of the Pad\'e poles, shown
in Figs. \ref{fig.v1n20} and \ref{fig.v1n30}, is almost symmetrical
with respect to the imaginary $k$-axis, despite the fact that we
fitted the $S$-matrix only on a short interval on the right hand side
of the real axis. This clearly shows that the approximate $S$-matrix
(\ref{productB}) when fitted to the exact points on that interval,
possesses the correct symmetry properties.

Comparing Figs. \ref{fig.v1n20} and \ref{fig.v1n30}, we see that, when
$N$ is increased, the meaningful poles converge to the exact ones
while the spurious poles change their positions. It is not difficult
to design an algorithm for automatically selecting the meaningful
poles. Indeed, as is seen, the meaningful are those poles that are
located in the fourth quadrant of the $k$-plane and do not have a
symmetric (or nearly-symmetric) partner at $-k$ (see
Eq. (\ref{spurious}) and the associated reasoning).

It should be noted that we did not make any effort to choose the best
fitting interval and the best distribution of the fitting points. This
is a model case, therefore, when doing the calculations, we already
knew the exact positions of the poles and could distribute the points
in such a way that the convergence would be much faster. In realistic
problems, however, no prior knowledge about resonance energies is
available. This is why we chose the fitting points in a sense
``arbitrarily''.  Despite this, the first few resonances were
reproduced to a high accuracy. In Table \ref{table.compareV1}, the
exact and approximate $S$-matrix poles are compared for the first five
resonances generated by the potential (\ref{pot1.formula}).

When applying this method to an unknown potential, one should do it in
a few iterations. At the first step the fitting interval and points
are chosen arbitrarily and the calculations are repeated with
different $N$. Looking at the results, it is easy to identify possible
meaningful poles. If an exact pole is very close to the real axis
(extremely narrow resonance) then the corresponding approximate pole
at this stage may be found slightly above the real axis. So, the poles
that are very close to the real axis but are located above it, should
not be excluded at the first iteration.

At the second step, one should choose a fitting interval that covers
all ``suspected'' resonances. If there is a possible narrow resonance,
the density of the fitting points should be higher next to
it. Repeating the calculations with the adjusted fitting interval, one
should be able to obtain the resonance energies and widths of at least
few most significant resonances.

A good guidance for choosing the fitting interval and fitting points
is the energy (or momentum) dependence of the scattering phase shift
$\delta_\ell(k)$, if available. Indeed, we are trying to construct
such $\tilde{S}_\ell(k)$ that would be as close as possible to the
exact function ${S}_\ell(k)=\exp[2i\delta_\ell(k)]$ on the real axis.
So, the fitting interval should cover all significant variations of
$\delta_\ell(k)$. The density of the fitting points should be high in
places where the function $\delta_\ell(k)$ changes very rapidly,
i.e. near narrow resonances.

\begin{figure}[ht!]
\centerline{\epsfig{file=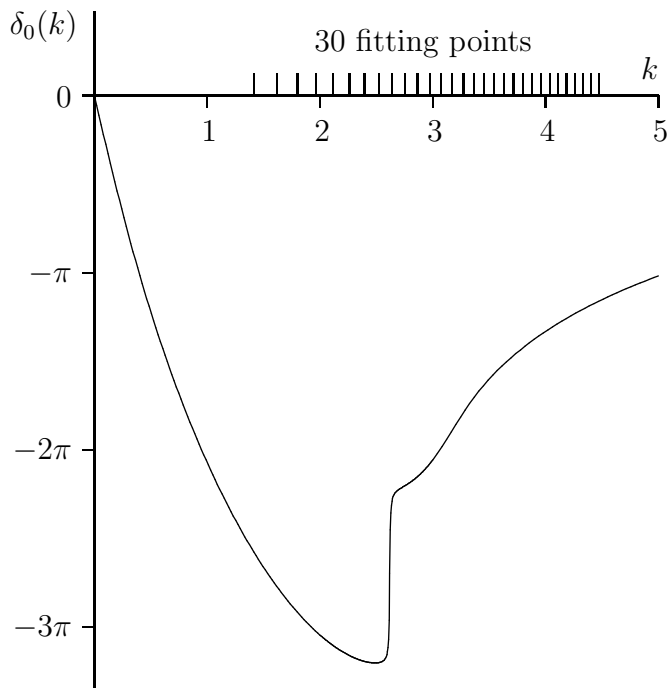}}
\caption{\sf
The exact $S$-wave phase-shift for the potential
(\protect\ref{pot1.formula}) as a function of the collision momentum.
The vertical bars on the $k$-axis show the points $k_n$, $n=1,2,\dots,
N$ ($N=30$) were
the exact and the approximate $S$-matrices coincide (the Pad\'e
fitting points).
}
\label{fig.deltaV1}
\end{figure}

\begin{figure}[ht!]
\centerline{\epsfig{file=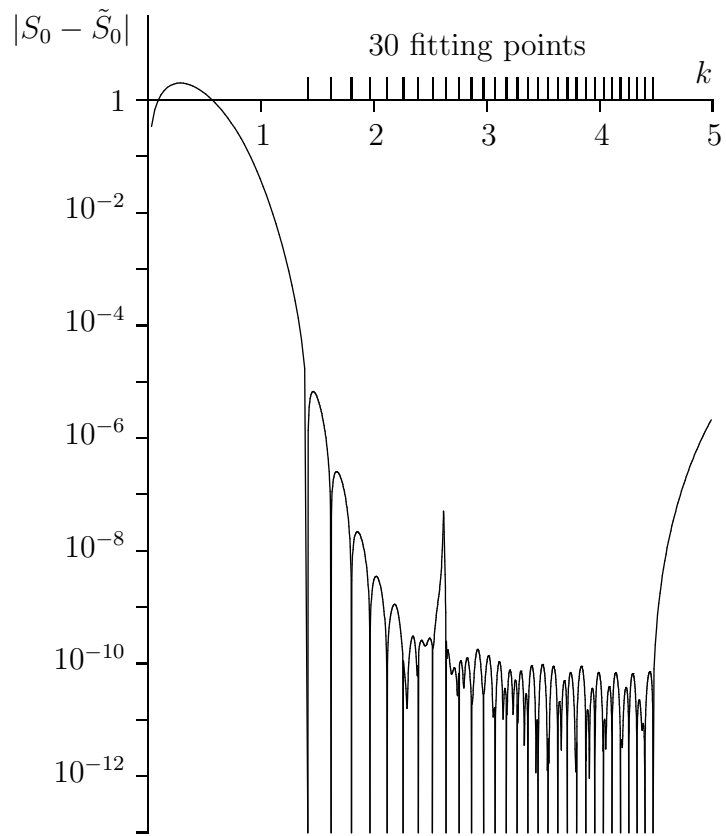}}
\caption{\sf
Error in the Pad\'e approximation of the $S$-matrix for the potential
(\protect\ref{pot1.formula}) with 30 points at which the error is zero.
}
\label{fig.sdiff}
\end{figure}

The $S$-wave phase shift function $\delta_0(k)$ for the potential
(\ref{pot1.formula}) together with our ``arbitrary'' choice of the
fitting points $k_1,k_2,\dots,k_N$ are shown in
Fig. \ref{fig.deltaV1}. As is seen, our choice of these points is not
the best. They should have been more dense where $\delta_0(k)$ jumps
in $\pi$ due to the first resonance. We however deliberately stick to
the simple set of points evenly distributed over the energy interval
$1\le E\le 10$. This is done for the purpose of testing the robustness
of the method with a poor choice of the fitting points.  The results
obtained show that even with such a choice the most significant
resonances are reproduced rather well.  The quality of the
approximation of the exact $S$-matrix on the fitting interval with
$N=30$ can be deduced from Fig. \ref{fig.sdiff}.

It may seem that one can avoid all this hassle with choosing the
fitting points by simply increasing $N$. Such an approach, however, is
not viable. The problem is that with large $N$ the linear system
(\ref{system_1}) becomes ill-conditioned because the neighboring
points $k_n$ and $k_{n+1}$ are too close to each other and the
corresponding equations of the linear system are only slightly
different. This is similar to a general and unavoidable problem of
fitting functions by high-order polynomials, where one has to solve
the so called Vandermonde system, which is ill-conditioned due to the
same reason (see Chapter 3 of Ref. \cite{num_recipes}). For the
potential (\ref{pot1.formula}), we found that the instabilities caused
by this problem start to show up when there are more than 70 points
placed on the interval $1\le E\le 10$.

\section{Conclusion}

The proposed method is based on a rigorous mathematical fact that
allows us to analytically continue the $S$-matrix, known on the real
axis, to the domains of complex momentum where it may have poles
corresponding to the bound and resonant states, which can thus be
easily located. Since the only input information we need, is a table
of the $S$-matrix values along the real $k$-axis, the method is
universal, i.e. independent of the nature of the underlying
interaction and the way the table is obtained.

The potential can be non-analytic that does not allow using the
complex rotation methods. It can be non-local or energy-dependent,
which also makes it extremely difficult to apply the rotation. And in
principle the $S$-matrix table can be obtained from experimental data
by means of the phase-shift analysis of the cross section. In all
these cases the spectral points can be immediately located as soon as
the $S$-matrix is calculated for real $k$.

The numerical examples show that the proposed method is stable and
accurate. With just few matching points, it reproduces the bound
states and the most significant resonances to the accuracy that is
sufficient for any practical purposes.


\newpage

\newpage
%
%
\begin{table}
\begin{center}
\begin{tabular}{|c|r|r|r|r|}
\hline
$N$ &
\multicolumn{1}{c|}{
${\rm Re\,}k\ ({\rm fm^{-1}})$} &
\multicolumn{1}{c|}{
${\rm Im\,}k\ ({\rm fm^{-1}})$} &
\multicolumn{1}{c|}{
${\rm Re\,}E\ ({\rm MeV})$} &
\multicolumn{1}{c|}{
${\rm Im\,}E\ ({\rm MeV})$}\\
\hline
\hline
1 &
$0.2\times10^{-16}$ &  $0.1595408980$ & {\boldmath $-1.0555482734$} &
$0.3\times10^{-15}$\\
\hline
\hline
2 &
$-0.6\times10^{-16}$ & $0.9148722594$ & $-34.7100271832$ &
$-0.5\times10^{-14}$\\
\cline{2-5}
&
$ 0.4\times10^{-16}$ & $0.2251639051$ & {\boldmath $ -2.1024785800$} &
$ 0.7\times10^{-15}$\\
\hline
\hline
3 &
$-0.1\times10^{-12}$ & $-6.5048412082$ & $-1754.71841570$ &
$0.7\times10^{-10}$\\
\cline{2-5}
&
$-0.9\times10^{-16}$ & $ 0.2305398780$ & {\boldmath $-2.2040739084$} &
$-0.2\times10^{-14}$\\
\cline{2-5}
&
$ 0.3\times10^{-14}$ & $ 0.7544907890$ & $-23.6070608641$ &
$ 0.2\times10^{-12}$\\
\hline
\hline
4 &
$-0.9\times10^{-12}$ & $ 1.4615496806$ & $-88.5852061340$ &
$-0.0000000001$\\
\cline{2-5}
&
$-0.2\times10^{-12}$ & $ 0.6623217412$ & $-18.1916485831$ &
$-0.9\times10^{-11}$\\
\cline{2-5}
&
$ 0.2\times10^{-14}$ & $ 0.2315901300$ & {\boldmath $ -2.2242014946$} &
$ 0.4\times10^{-13}$\\
\cline{2-5}
&
$ 0.8\times10^{-12}$ & $-0.9942646075$ & $-40.9956706906$ &
$-0.6\times10^{-10}$\\
\hline
\hline
5 &
$-0.2\times10^{-10}$ & $ 1.3860254952$ & $-79.6666349435$ &
$-0.0000000021$\\
\cline{2-5}
&
$-0.1\times10^{-11}$ & $ 0.6586014432$ & $-17.9878555574$ &
$-0.6\times10^{-10}$\\
\cline{2-5}
&
$ 0.4\times10^{-14}$ & $ 0.2316032946$ & {\boldmath $-2.2244543701$} &
$ 0.8\times10^{-13}$\\
\cline{2-5}
&
$ 0.8\times10^{-11}$ & $-0.9618425253$ & $-38.3655990692$ &
$-0.0000000007$\\
\cline{2-5}
&
$0.0000000207$       & $-85.5633495380$& $-303605.468939$ &
$-0.0001469872$\\
\hline
\end{tabular}
\caption{
Poles of the approximate $S$-wave function $\tilde{S}_0(k)$, found for
the potential (\protect\ref{pot.NN1})
with $N$ fitting points $k_1,k_2,\dots,k_N$ evenly distributed over
the interval $1\,{\rm MeV}\le E\le 10\,{\rm MeV}$. For $N=1$,
the single fitting point corresponds to $E=1\,{\rm MeV}$.
The boldfaced numbers converge to the exact bound-state energy
$-2.2244674752$\,MeV.
}
\label{table.nn1}
\end{center}
\end{table}
\begin{table}
\begin{center}
\begin{tabular}{|c|r|r|r|r|}
\hline
$N$ &
\multicolumn{1}{c|}{
${\rm Re\,}k\ ({\rm fm^{-1}})$} &
\multicolumn{1}{c|}{
${\rm Im\,}k\ ({\rm fm^{-1}})$} &
\multicolumn{1}{c|}{
${\rm Re\,}E\ ({\rm MeV})$} &
\multicolumn{1}{c|}{
${\rm Im\,}E\ ({\rm MeV})$}\\
\hline
\hline
1 &
$-0.2\times10^{-16}$ & $-0.0753661730$ & {\boldmath $-0.2355520897$} &
$ 0.1\times10^{-15}$\\
\hline
\hline
2 &
$-0.8\times10^{-16}$ & $ 0.7896265828$ & $-25.8569655146$ &
$-0.5\times10^{-14}$\\
\cline{2-5}
&
$-0.1\times10^{-16}$ & $-0.0409220958$ & {\boldmath $-0.0694464054$} &
$ 0.4\times10^{-16}$\\
\hline
\hline
3 &
$-0.2\times10^{-14}$ & $ 0.7187567981$ & $-21.4238720550$ &
$-0.1\times10^{-12}$\\
\cline{2-5}
&
$ 0.2\times10^{-16}$ & $-0.0399961842$ & {\boldmath $-0.0663393412$} &
$-0.8\times10^{-16}$\\
\cline{2-5}
&
$ 0.4\times10^{-12}$ & $-11.9310494904$ & $-5903.25209234$ &
$-0.0000000004$\\
\hline
\hline
4 &
$-0.9\times10^{-12}$ & $-1.1951630087$ & $-59.2363541815$ &
$0.8\times10^{-10}$\\
\cline{2-5}
&
$-0.2\times10^{-15}$ & $-0.0399135508$ & {\boldmath $-0.0660655059$} &
$0.7\times10^{-15}$\\
\cline{2-5}
&
$0.6\times10^{-13}$ & $ 0.6857755943$ & $-19.5028502331$ &
$0.4\times10^{-11}$\\
\cline{2-5}
&
$0.1\times10^{-11}$ & $ 1.4486760697$ & $-87.0315278526$ &
$0.0000000001$\\
\hline
\hline
5 &
$-0.0000005377$ & $-259.016625071$ & $-2782206.21227$ &
$0.0115523186$\\
\cline{2-5}
&
$-0.5\times10^{-10}$ & $-1.1697429680$ & $-56.7433434047$ &
$0.0000000053$\\
\cline{2-5}
&
$-0.7\times10^{-15}$ & $-0.0399132475$ & {\boldmath $-0.0660645021$} &
$0.2\times10^{-14}$\\
\cline{2-5}
&
$0.2\times10^{-11}$ & $0.6850556465$ & $-19.4619223746$ &
$0.9\times10^{-10}$\\
\cline{2-5}
&
$0.9\times10^{-10}$ & $1.4068361606$ & $-82.0769256396$ &
$0.0000000102$\\
\hline
\end{tabular}
\caption{
Poles of the approximate $S$-wave function $\tilde{S}_0(k)$, found for
the potential (\protect\ref{pot.NN0})
with $N$ fitting points $k_1,k_2,\dots,k_N$ evenly distributed over
the interval $1\,{\rm MeV}\le E\le 10\,{\rm MeV}$. For $N=1$,
the single fitting point corresponds to $E=1\,{\rm MeV}$.
The boldfaced numbers converge to the exact virtual-state energy
$-0.0660644719$\,MeV.
}
\label{table.nn0}
\end{center}
\end{table}
\begin{table}
\begin{center}
\begin{tabular}{|c|c|c|c|}
\hline
${\rm Re\,}k$ & ${\rm Im\,}k$ & $E_{\rm r}$ & $\Gamma$\\
\hline
2.6177861703 & $-0.0048798793$ & 3.4263903101 & 0.0255489612\\
\hline
3.1300424437 & $-0.3571442525$ & 4.8348068411 & 2.2357533377\\
\hline
3.3983924252 & $-0.9972518977$ & 5.2772798640 & 6.7781065905\\
\hline
3.5914630921 & $-1.6639555063$ & 5.0649296074 & 11.9520695757\\
\hline
3.7383048831 & $-2.3317810362$ & 4.2688602993 & 17.4338168679\\
\hline
3.8534573944 & $-2.9922517758$ & 2.9477816003 & 23.0610294625\\
\hline
3.9448589582 & $-3.6424641005$ & 1.1471837383 & 28.7380142741\\
\hline
4.0173695706 & $-4.2816627873$ & $-1.0966889789$ & 34.4020435866\\
\hline
4.0742577101 & $-4.9099759809$ & $-3.7541441225$ & 40.0090149934\\
\hline
\end{tabular}
\caption{\sf
The $S$-wave resonance points, $k^2/(2m)=E_{\rm r}-i\Gamma/2$,
of the exact $S$-matrix for the potential
(\protect\ref{pot1.formula}). All the values are given in the
arbitrary units such that $\hbar=m=1$. They were obtained using the
Jost function method described in Ref. \protect\cite{exactmethod}.
}
\label{table.V1exact}
\end{center}
\end{table}
\begin{table}
\begin{center}
\begin{tabular}{|l|c|c|}
\hline
& ${\rm Re\,}k$ & ${\rm Im\,}k$\\
\hline
exact       & 2.6177861703 & $-0.0048798793$\\
approximate & 2.6177861702 & $-0.0048798793$\\
\hline
exact       & 3.1300424437 & $-0.3571442525$\\
approximate & 3.1300424420 & $-0.3571442539$\\
\hline
exact       & 3.3983924252 & $-0.9972518977$\\
approximate & 3.3984191164 & $-0.9972641751$\\
\hline
exact       & 3.5914630921 & $-1.6639555063$\\
approximate & 3.6001916924 & $-1.6689691843$\\
\hline
exact       & 3.7383048831 & $-2.3317810362$\\
approximate & 3.8771630790 & $-2.5129915936$\\
\hline
\end{tabular}
\caption{\sf
Comparison of the first five resonance points for the
potential (\protect\ref{pot1.formula}), obtained using the rigorous Jost
function method~\protect\cite{exactmethod} (exact) and the Pad\'e
approximation with the number of fitting points $N=30$ evenly distributed over
the interval $1\le E\le 10$ (the units are such that $\hbar=m=1$).
}
\label{table.compareV1}
\end{center}
\end{table}
\end{document}